\documentclass[letterpaper,10pt]{article} 
\usepackage{amsmath,amsthm,amssymb,amsfonts}
\usepackage{graphicx,float}

\usepackage{amsmath}
\usepackage{amssymb}
\usepackage{graphicx}
\usepackage{graphicx}
\usepackage{dcolumn}
\usepackage{bm}
\usepackage{amsfonts}
\usepackage{latexsym}
\usepackage{pdfsync}
    \usepackage{fullpage} 
\usepackage{colordvi}
\usepackage{color}

\begin{document}
\newcommand{\commentout}[1]{}

\newcommand{\nwc}{\newcommand}
\newcommand{\bz}{{\mathbf z}}
\newcommand{\sqk}{\sqrt{\ks}}
\newcommand{\sqkone}{\sqrt{|\ks_1|}}
\newcommand{\sqktwo}{\sqrt{|\ks_2|}}
\newcommand{\invsqkone}{|\ks_1|^{-1/2}}
\newcommand{\invsqktwo}{|\ks_2|^{-1/2}}
\newcommand{\partz}{\frac{\partial}{\partial z}}
\newcommand{\grady}{\nabla_{\by}}
\newcommand{\gradp}{\nabla_{\bp}}
\newcommand{\invf}{\cF^{-1}_2}
\newcommand{\myphi}{\Phi_{(\eta,\rho)}}
\newcommand{\minrg}{|\min{(\rho,\gamma^{-1})}|}
\newcommand{\al}{\alpha}
\newcommand{\xvec}{\vec{\mathbf x}}
\newcommand{\kvec}{{\vec{\mathbf k}}}
\newcommand{\lt}{\left}
\newcommand{\ksq}{\sqrt{\ks}}
\newcommand{\rt}{\right}
\nwc{\bG}{{\bf G}}
\newcommand{\ga}{\gamma}
\newcommand{\vas}{\varepsilon}
\newcommand{\lan}{\left\langle}
\newcommand{\ran}{\right\rangle}
\newcommand{\tvas}{{W_z^\vas}}
\newcommand{\psiep}{{W_z^\vas}}
\newcommand{\wep}{{W^\vas}}
\newcommand{\weptil}{{\tilde{W}^\vas}}
\newcommand{\wepz}{{W_z^\vas}}
\newcommand{\weps}{{W_s^\ep}}
\newcommand{\wepsp}{{W_s^{\ep'}}}
\newcommand{\wepzp}{{W_z^{\vas'}}}
\newcommand{\wepztil}{{\tilde{W}_z^\vas}}
\newcommand{\vvas}{{\tilde{\ml L}_z^\vas}}
\newcommand{\veptil}{{\tilde{\ml L}_z^\vas}}
\newcommand{\vep}{{{ V}_z^\vas}}
\newcommand{\cvc}{{{\ml L}^{\ep*}_z}}
\newcommand{\cvcp}{{{\ml L}^{\ep*'}_z}}
\newcommand{\cvp}{{{\ml L}^{\ep*'}_z}}
\newcommand{\cvtil}{{\tilde{\ml L}^{\ep*}_z}}
\newcommand{\cvtilp}{{\tilde{\ml L}^{\ep*'}_z}}
\newcommand{\vtil}{{\tilde{V}^\ep_z}}
\newcommand{\ktil}{\tilde{K}}
\newcommand{\n}{\nabla}
\newcommand{\tkappa}{\tilde\kappa}
\newcommand{\ks}{{\omega}}
\newcommand{\mbx}{\mb x}
\newcommand{\br}{\mb r}
\nwc{\bH}{{\mb H}}
\newcommand{\bu}{\mathbf u}
\nwc{\bxp}{{{\mathbf x}}}
\nwc{\byp}{{{\mathbf y}}}
\newcommand{\bD}{\mathbf D}
\nwc{\bh}{\mathbf h}
\newcommand{\bB}{\mathbf B}
\newcommand{\bb}{\mathbf b}
\newcommand{\bC}{\mathbf C}
\nwc{\cO}{\mathcal  O}
\newcommand{\bp}{\mathbf p}
\newcommand{\bq}{\mathbf q}
\newcommand{\by}{\mathbf y}
\nwc{\bP}{\mathbf P}
\nwc{\bs}{\mathbf s}
\nwc{\bX}{\mathbf X}
\newcommand{\pdg}{\bp\cdot\nabla}
\newcommand{\pdgx}{\bp\cdot\nabla_\bx}
\newcommand{\one}{1\hspace{-4.4pt}1}
\newcommand{\corr}{r_{\eta,\rho}}
\newcommand{\rinf}{r_{\eta,\infty}}
\newcommand{\rzero}{r_{0,\rho}}
\newcommand{\rzeroinf}{r_{0,\infty}}
\nwc{\om}{\omega}
\nwc{\Gp}{{G_{\rm par}}}
\nwc{\nwt}{\newtheorem}
\nwc{\xp}{{x^{\perp}}}
\nwc{\yp}{{y^{\perp}}}
\nwt{remark}{Remark}
\nwt{definition}{Definition} 
\nwc{\bd}{{\mb d}}
\nwc{\ba}{{\mb a}}
\nwc{\mbe}{{\mb e}}
\nwc{\bal}{\begin{align}}
\nwc{\bea}{\begin{eqnarray}}
\nwc{\beq}{\begin{eqnarray}}
\nwc{\bean}{\begin{eqnarray*}}
\nwc{\beqn}{\begin{eqnarray*}}
\nwc{\beqast}{\begin{eqnarray*}}

\nwc{\eal}{\end{align}}
\nwc{\eea}{\end{eqnarray}}
\nwc{\eeq}{\end{eqnarray}}
\nwc{\eean}{\end{eqnarray*}}
\nwc{\eeqn}{\end{eqnarray*}}
\nwc{\eeqast}{\end{eqnarray*}}

\nwc{\ep}{\varepsilon}
\nwc{\eps}{\varepsilon}
\nwc{\ept}{\epsilon}
\nwc{\vrho}{\varrho}
\nwc{\orho}{\bar\varrho}
\nwc{\ou}{\bar u}
\nwc{\vpsi}{\varpsi}
\nwc{\lamb}{\lambda}
\nwc{\Var}{{\rm Var}}

\nwt{cor}{Corollary}
\nwt{proposition}{Proposition}
\nwt{corollary}{Corollary}
\nwt{theorem}{Theorem}
\nwt{summary}{Summary}
\nwt{lemma}{Lemma}
\nwc{\nn}{\nonumber}
\nwc{\mf}{\mathbf}
\nwc{\mb}{\mathbf}
\nwc{\ml}{\mathcal}
\nwc{\bj}{{\mb j}}
\nwc{\bA}{{\mb A}}
\nwc{\IA}{\mathbb{A}} 
\nwc{\bi}{\mathbf i}
\nwc{\bo}{\mathbf o}
\nwc{\IB}{\mathbb{B}}
\nwc{\IC}{\mathbb{C}} 
\nwc{\ID}{\mathbb{D}} 
\nwc{\IM}{\mathbb{M}} 
\nwc{\IP}{\mathbb{P}} 
\nwc{\bI}{\mathbf{I}} 
\nwc{\IE}{\mathbb{E}} 
\nwc{\IF}{\mathbb{F}} 
\nwc{\IG}{\mathbb{G}} 
\nwc{\IN}{\mathbb{N}} 
\nwc{\IQ}{\mathbb{Q}} 
\nwc{\IR}{\mathbb{R}} 
\nwc{\IT}{\mathbb{T}} 
\nwc{\IZ}{\mathbb{Z}} 
\nwc{\II}{\mathbb{I}} 

\nwc{\cE}{{\ml E}}
\nwc{\cP}{{\ml P}}
\nwc{\cQ}{{\ml Q}}
\nwc{\cL}{{\ml L}}
\nwc{\cX}{{\ml X}}
\nwc{\cW}{{\ml W}}
\nwc{\cZ}{{\ml Z}}
\nwc{\cR}{{\ml R}}
\nwc{\cV}{{\ml V}}
\nwc{\cT}{{\ml T}}
\nwc{\crV}{{\ml L}_{(\delta,\rho)}}
\nwc{\cC}{{\ml C}}
\nwc{\cA}{{\ml A}}
\nwc{\cK}{{\ml K}}
\nwc{\cB}{{\ml B}}
\nwc{\cD}{{\ml D}}
\nwc{\cF}{{\ml F}}
\nwc{\cS}{{\ml S}}
\nwc{\cM}{{\ml M}}
\nwc{\cG}{{\ml G}}
\nwc{\cH}{{\ml H}}
\nwc{\cN}{{\ml N}}
\nwc{\bk}{{\mb k}}
\nwc{\bT}{{\mb T}}
\nwc{\cbz}{\overline{\cB}_z}
\nwc{\supp}{{\hbox{\rm supp}}}
\nwc{\fR}{\mathfrak{R}}
\nwc{\bY}{\mathbf Y}
\newcommand{\mbr}{\mb r}
\nwc{\pft}{\cF^{-1}_2}
\nwc{\bU}{{\mb U}}
\nwc{\bPhi}{{\mb \Phi}}
\nwc{\bPsi}{{\mb \Psi}}

\commentout{
\newenvironment{proof}[1][Proof]{\begin{trivlist}
\item[\hskip \labelsep {\bfseries #1}]}{\end{trivlist}}
\newenvironment{definition}[1][Definition]{\begin{trivlist}
\item[\hskip \labelsep {\bfseries #1}]}{\end{trivlist}}
\newenvironment{example}[1][Example]{\begin{trivlist}
\item[\hskip \labelsep {\bfseries #1}]}{\end{trivlist}}
\newenvironment{remark}[1][Remark]{\begin{trivlist}
\item[\hskip \labelsep {\bfseries #1}]}{\end{trivlist}}
}

\newcommand{\CC}{\mathbb{C}}
\newcommand{\hj}{\hat{J}}
\newcommand{\tj}{\tilde{J}}
\newcommand{\xmax}{x_{\text{max}}}
\newcommand{\xmin}{x_{\text{min}}}
\newcommand{\bx}{{x}}
\newcommand{\be}{\bar{e}}
\newcommand{\emax}{c_{\text{max}}}
\nwc{\red}{\color{red}}
\nwc{\blue}{\color{blue}}
\nwc{\green}{\color{green}}
\newcommand{\RR}{\mathbb{R}}
\nwc{\suppx}{{\hbox{supp}(\mathbf x)}}
\nwc{\bn}{\mathbf{n}}

\title{\bf Mismatch and resolution in compressive imaging }
\author{Albert Fannjiang\thanks{Corresponding author:  fannjiang@math.ucdavis.edu. 
} \quad  Wenjing Liao\\ \\
   Department of Mathematics,
    University of California, Davis, CA 95616-8633.    }
\date{}
\maketitle
\thispagestyle{empty}
\centerline{\bf \small ABSTRACT}

\bigskip

Highly coherent sensing matrices arise  in discretization of 
continuum problems 
such as radar  and medical imaging when the grid spacing
is below the Rayleigh threshold as well as in using
highly coherent, redundant dictionaries as sparsifying operators.

Algorithms (BOMP, BLOOMP)  based on techniques of band exclusion and local optimization  are proposed to enhance Orthogonal Matching
Pursuit (OMP) and deal with such coherent sensing matrices. 

BOMP and BLOOMP have provably performance guarantee of 
reconstructing sparse, widely separated objects 
{\em independent}  of the redundancy and
 have a sparsity constraint and computational
cost similar to OMP's.

Numerical study  demonstrates  the effectiveness  of BLOOMP for
compressed sensing  with  highly coherent, redundant sensing matrices. \\


\bigskip

\centerline{\bf \small 1. INTRODUCTION}

\bigskip

Model mismatch is a fundamental issue in imaging and image processing$^3$. To reduce mismatch error, it is often
necessary to consider measurement matrices that are highly
coherent and redundant. 
Such  measurement matrices  lead to
serious difficulty in applying compressed sensing (CS) techniques.

Let us consider two examples:
discretization in analog imaging and 
sparse representation of signals. 

\begin{figure}
\begin{center}
\includegraphics[width=10cm]{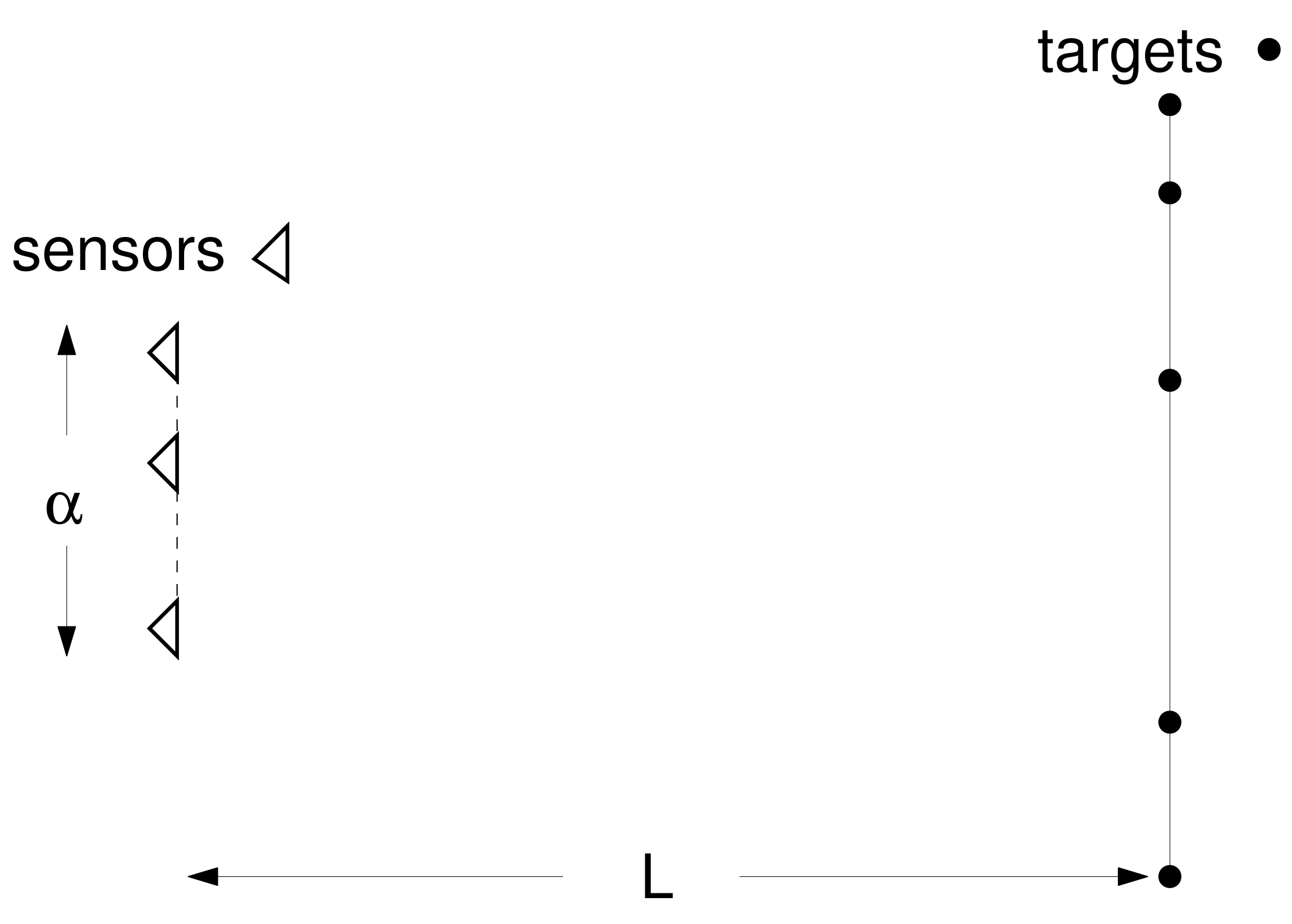}
\caption{Sensors are distributed in an  aperture of linear size $A$ on the left ($z=0$)  and the point sources of unknown locations are distributed
on the target plane on the right ($z=L$).}
\label{fig0}
\end{center}
\end{figure}

Consider remote sensing of point sources as depicted in figure
\ref{fig0}.  Let the noiseless signal at the point $r$ on the sensor plane  emitted by  the unit source at $\xi$ on 
the target plane be given by the paraxial Green function 
\beq
G(r,  \xi)&=&{e^{i\om L}\over 4\pi L} \times { \exp{\lt(i\om|r-\xi|^2\over 2L\rt)} }\nn\\
&=&{e^{i\om L}\over 4\pi L} \exp{\lt(i\om r^2\over 2L\rt)}  { \exp{\lt(-i\om r\xi\over L\rt)}}\exp{\lt(i\om \xi ^2\over 2L\rt)} \label{8-3}
\eeq
where $\om$ is the wavenumber. 
Suppose that $s$ point sources of unknown locations $\xi_j$
and strengths $c_j, j=1,...,s$
emit simultaneously. 
Then 
the signals received by the sensors $l, l=1,...,N$ are
\beq
\label{40}\label{50}
y_l=\sum_{j=1}^s c_j G(r_l, \xi_j)+n_l,\quad l=1,\ldots, N 
\eeq
where $n_l$ are external noise. 
 
To cast eq. (\ref{40}) in the form of finite, discrete linear inversion problem
let 
 $\cG=\{p_1,\ldots, p_M\}$  be a regular grid of spacing $\ell$
smaller than the minimum distance among the targets.
 Consequently, each grid point has {\em at most}  one target within
 the distance $\ell/2$.  
 Write $\mbx=(x_j)\in \IC^M$ with
\[
x_j= \exp{\lt(i\om p_j ^2\over 2L\rt)}c_{j'}
\]
whenever $p_j$ is within $\ell/2$ from  
some target $j'$   and zero otherwise. 
When a target is located at the midpoint between two neighboring grid points,
we can associate either grid point with the target. 

Let the data vector $\bb=(b_l)\in \IC^N$ be defined as
\beq
b_l=N^{-1/2}{4\pi L e^{-i\om L}} e^{-i\om r^2\over 2L} y_l
\eeq
and the measurement matrix be 
\beq
\label{2}
\bA=\begin{bmatrix}
     \ba_{1} & \ldots & \ba_{M}
  \end{bmatrix}\in \IC^{N\times M}
  \eeq
  with
  \beq
\label{3}
 \ba_j={1\over \sqrt{N}}\lt(  { \exp{\lt(-i\om r_k p_j \over L\rt)}} \rt) \in \IC^N,\quad j=1,...,M.
  \eeq
   After proper normalization of noise we 
rewrite the problem in the form
\beq\label{1}
    \bA \mbx + \mbe = \bb
    \label{linearsystem}
\eeq
where the error vector 
 $\mbe=(e_k)\in \IC^N$ is the sum of  the external noise
 $\bn=(n(t_k))$ and the discretization or gridding  error $\bd=(\delta_k) \in \IC^N $
 due to approximating 
 the locations  by the grid points in $\cG$.  
Obviously  the discretization error decreases as
the grid spacing $\ell$ decreases.  The discretization error, however,  depends {\em nonlinearly} on the objects and hence is not in the form of either additive or multiplicative noise.

We shall consider in this paper only  random sampling over an aperture $\alpha$ satisfying 
 the Rayleigh criterion$^1$ 
 \beq
 \label{Ray}
 \alpha\geq { L\lambda\over \ell}
 \eeq
 where $\lambda=2\pi/\om$ is the wavelength. This sets the limit
 of the resolution
 \beq
 \ell\geq {L\lambda\over \alpha}\equiv \ell_R
 \eeq
whose right hand side shall be referred to as the Rayleigh length (RL). 
 
To reduce the gridding error,  consider 
the fractional  grid  with spacing 
\beq
\label{23}
\ell=\ell_R/F
\eeq
for
some large integer $F\in \IN$ called  {\em the
refinement factor}. The
relative gridding error $\|\bd\|_2/\|\bb\|_2$ is roughly inversely proportional to the refinement factor.

On the other hand, a large refinement factor leads to difficulty in applying compressed sensing techniques. 
 A practical indicator of the CS performance  is the mutual coherence \begin{equation}
  \mu(\bA)= \max_{k\neq l} {|\lan \ba_{k},\ba_{l}\ran |\over |\ba_k| |\ba_l|}, 
  \label{innerproduct}
\end{equation}
which  increases 
 with $F$ as  the near-by columns of the sensing matrix
become highly correlated.
Indeed, for $F=1$ ,
$\mu(\bA)$ decays like  
 $\cO(N^{-1/2})$ while for $F>1$ $\mu(\bA)=\cO(1)$. 


\begin{figure}[t]
\centering
  \includegraphics[width=8cm]{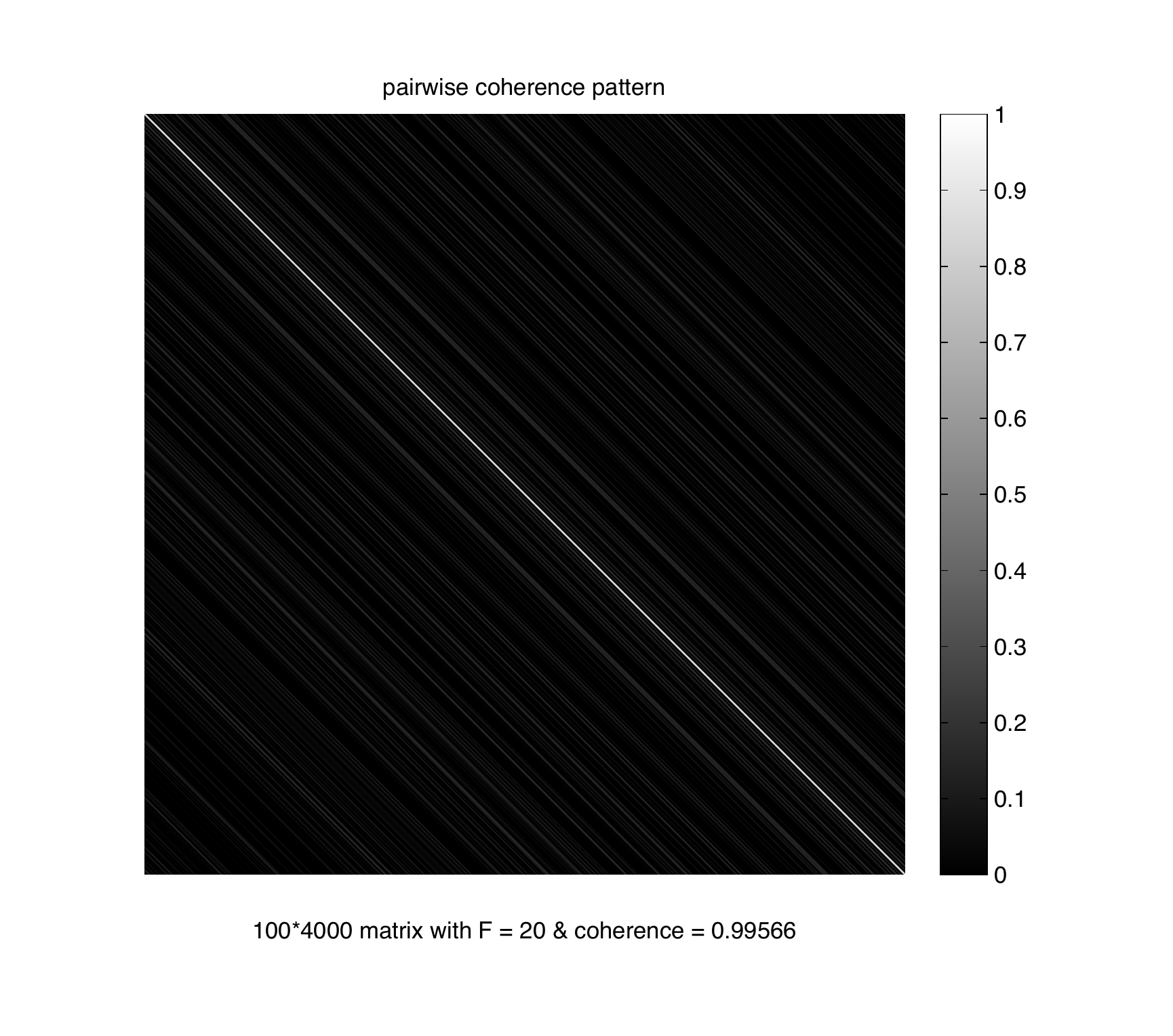}
   \includegraphics[width=8cm,height=7cm]{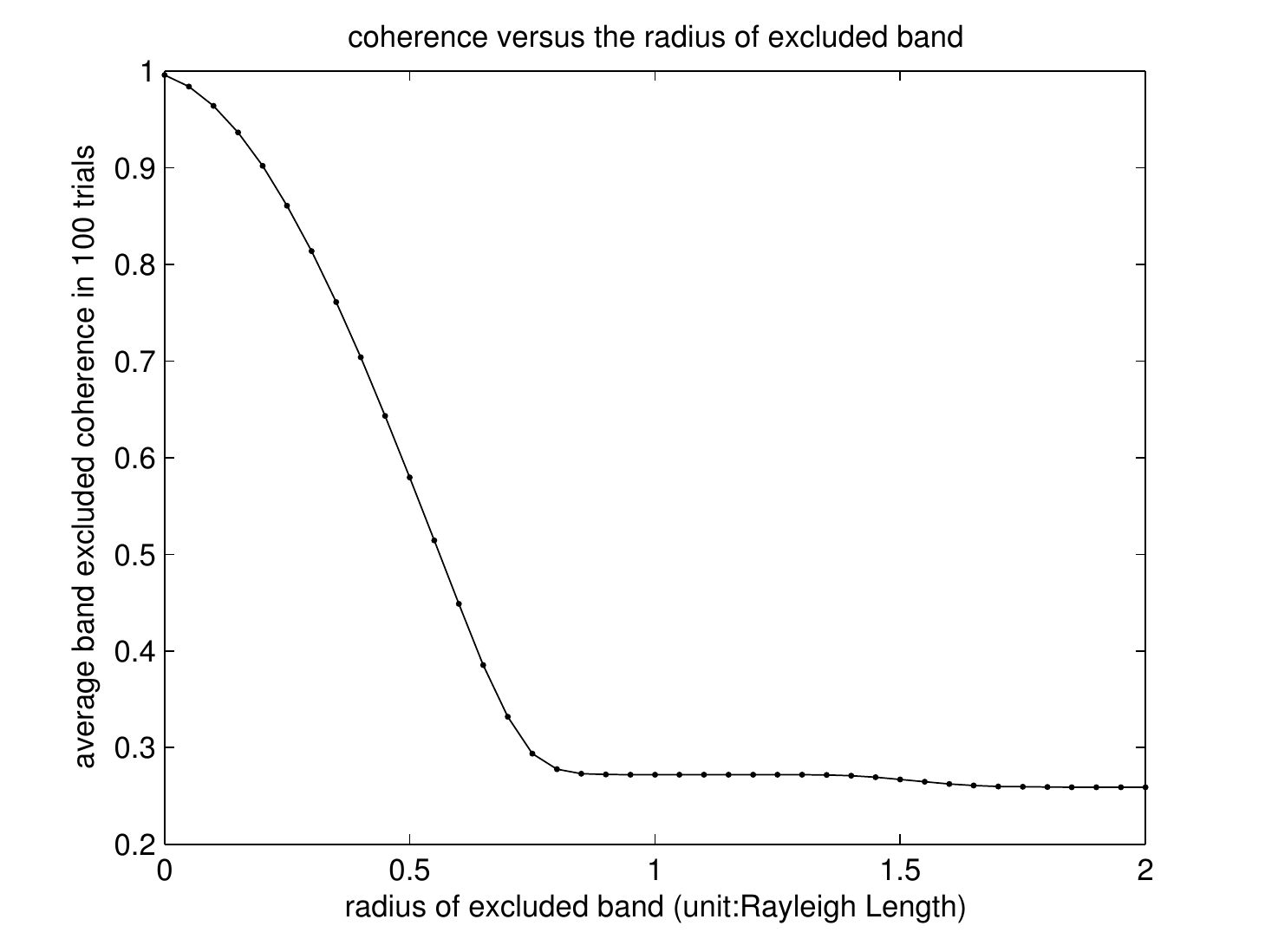}
 \caption{Left: The  coherence pattern $[\mu(j,k)]$ of
a $100\times 4000$ matrix (\ref{3}) 
with $F=20$. Right:  A semi-cross section of the coherence band averaged over 100 independent realizations of (\ref{2}).  }
    \label{fig1}
\end{figure}
Figure \ref{fig1} shows 
the coherence pattern from the  one-dimensional setting.  In two or three dimensions, the
coherent pattern is more complicated because the coherence band
corresponds to the higher dimensional neighborhood.  

\commentout{
\begin{figure}
\centering
\includegraphics[width=7.5cm]{figures/DFFCoherenceVersusBand.eps}
\includegraphics[width=7.5cm]{figures/GaussianCOherenceVersusBand.eps}
\caption{The coherence bands of the DFT frame $\bPsi$ (left) and $\bA=\bPhi\bPsi$ (right), the latter being averaged over
100 trials. 
}
\label{fig10}
\end{figure}
}

 More generally, coherent bands can arise
in sparse and redundant  representation by  overcomplete dictionaries. 
Following Duarte and Baraniuk$^5$  we consider the following CS problem  
\beq
\label{101}
\bb= \bPhi \by + \mbe
\eeq
with a $N \times R$ i.i.d Gaussian matrix $\bPhi$
where 
the signal $\by$  is represented by a redundant dictionary $\bPsi$.
For example, suppose the sparsifying dictionary is
the over-sampled, redundant DFT frame 
\beq
\label{102}
\Psi_{k,j} = \frac{1}{\sqrt{R}} e^{-2\pi i \frac{(k-1)(j-1)}{RF}},\quad k=1,...,R,\quad j=1,...,RF.
\eeq
where  $F$ is  the redundancy factor. Writing $\by =\bPsi \mbx$  we have the same  form (\ref{linearsystem}) with $
\bA=\bPhi\bPsi
$. The  coherence bands of $\bPsi$ and $\bA$  have the similar structure as shown in Figure \ref{fig1}.

Without extra prior information besides the object sparsity,
the CS techniques can not guarantee to  recover the objects.
The additional prior information we impose here is
that  the objects
are sufficiently  separated with respect to 
the coherence band (see below for details). And 
 we propose  modified versions  of Orthogonal Matching Pursuit (OMP) 
for handling  highly coherent measurement matrices. 

The rest of the paper is organized as follows. In Section 2, we introduce
the algorithm, BOMP, to deal with highly coherent measurement matrices and states a performance guarantee for BOMP. In Section 3, 
we introduce another technique, Local Optimization, to enhance BOMP's
performance and state the performance guarantee for the resulting algorithm, BLOOMP.  In Section 4, we present  numerical results
for the two examples discussed above and compare the existing 
algorithms with ours. We conclude in Section 5.  \\ 

\bigskip

\centerline{\bf\small 2. BAND-EXCLUDED OMP (BOMP)}

\bigskip

The first technique that we introduce to take advantage of the
prior information of widely separated objects  is called Band Exclusion and 
can be easily embedded in  the greedy algorithm,
Orthogonal Matching Pursuit (OMP). 

 Let $\eta>0$. Define the $\eta$-coherence band  of the index $k$ as \begin{equation}
   B_\eta(k) = \{i\ | \ \mu(i,k) > \eta\},
   \label{singleneighbor}
\end{equation}
and   the secondary coherence band as
\beq
   B^{(2)}_\eta(k) &\equiv& B_\eta(B_\eta(k))= \displaystyle \cup_{j\in B_\eta(k)} B_\eta(j)
\eeq

Embedding the technique of coherence band exclusion in  OMP
yields the following algorithm.

\bigskip

\begin{center}
   \begin{tabular}{l}
   \hline
   
   \centerline{{\bf Algorithm 1.}\quad Band-excluded Orthogonal Matching Pursuit (BOMP)} \\ \hline
   Input: $\bA, \bb,\eta>0$\\
 Initialization:  $\mbx^0 = 0, \br^0 = \bb$ and $S^0=\emptyset$ \\ 
Iteration: For  $n=1,...,s$\\
\quad {1) $i_{\rm max} = \hbox{arg}\max_{i}|\lan \br^{n-1},\ba_i\ran | , i \notin B^{(2)}_\eta(S^{n-1}) $} \\
  \quad      2) $S^{n} = S^{n-1} \cup \{i_{\rm max}\}$ \\
  \quad  3) $\mbx^n = \hbox{arg} \min_\bz \|
     \bA \bz-\bb\|_2$ s.t. \hbox{supp}($\bz$) $\in S^n$ \\
  \quad   4) $\br^n = \bb- \bA \mbx^n$\\
 Output: $\mbx^s$. \\
 \hline
   \end{tabular}
\end{center}

\bigskip

 We have 
  the following  performance guarantee for BOMP$^8$. 
\begin{theorem} 
\label{thm1}

Let $\mbx$ be $s$-sparse. Let $\eta>0$ be fixed. 
Suppose that
\beq
\label{sep}
B_\eta(i)\cap B^{(2)}_\eta(j)=\emptyset, \quad \forall i, j\in \hbox{supp}(\mbx)
\eeq
and that
    \beq
  \eta(5s -4)\frac{\xmax}{\xmin} + \frac{5\|\mbe\|_2}{2\xmin} < 1 
    \label{RMIP}
    \eeq
    where
    \[
    \xmax = \max_{k} |x_k|,\quad  \xmin = \min_{k} |x_k|.
    \]
    Let $\hat\mbx$ be the BOMP reconstruction. 
Then $\hbox{supp}(\hat\mbx)\subseteq B_\eta(\hbox{supp}(\mbx))$
and moreover every nonzero component of $\hat\mbx$ is in
the $\eta$-coherence band of a unique nonzero component of $\mbx$. 
\end{theorem}

\begin{remark}
\label{rmk1}

In the case of the matrix 
(\ref{3}),  if every two indices in $\suppx$ is more than
one RL apart, then $\eta$ is small for sufficiently
large $N$, cf. Figure ~$\ref{fig1}$. 
  
 When the dynamic range ${\xmax}/{\xmin} = \cO(1)$, 
 Theorem \ref{thm1} guarantees approximate recovery
 of  $\cO(\eta^{-1})$ sparsity pattern by BOMP. Since
 $\eta=\cO(N^{-1/2})$ for $N\gg 1$, the sparsity constrain by
 (\ref{RMIP}) has the same order of magnitude as 
 the condition for OMP's performance$^4$ in the presence of noise.

The main difference between (\ref{RMIP})  and the OMP result
 lies in the role
played by  the dynamic range $\xmax/\xmin$ which is absent
in the condition for OMP's performance.
Numerical evidence points  to the sensitive dependence of BOMP's performance on dynamic range (Figure \ref{fig5}). 
\end{remark}

\begin{remark}
\label{rmk2}
Condition  (\ref{sep}) means 
that BOMP has a resolution length no worse than 3 $\ell_R$ {\em independent} of the refinement factor.
Numerical experiments show that 
BOMP can resolve 
 objects separated by close to 1  $\ell_R$  when the dynamic
 range is close to 1.

\end{remark}

\bigskip

\centerline{\bf \small 3. BAND-EXCLUDED, LOCALLY OPTIMIZED OMP (BLOOMP)}

\bigskip

We now introduce 
the second technique, 
the {\em Local Optimization} (LO), to improve the performance of
BOMP.  

LO is a residual-reduction technique  applied
to the current estimate $S^k$ of the object support.  
To this end,  we minimize  the residual  ${\|\bA \hat\mbx-\bb\|_2}$ by varying  one location at a time 
while all other locations held fixed.
In each step we consider  $\hat\mbx$ whose support
differs from $S^n$ by at most one index  in the  coherence band of $S^n$ but whose amplitude is chosen to minimize
the residual. The search is local in the sense that
  during the search in the coherence band of one nonzero component
the locations of other nonzero components are fixed.
The total number of search is $\cO(s^2F)$. 
The amplitudes of the improved estimate  is carried out by solving the least squares problem. Because of
the local nature of the LO step, the computation is not
expensive. 

\bigskip

\begin{center}
   \begin{tabular}{l}
   \hline  
   \centerline{{\bf Algorithm 2.}\quad  Local Optimization (LO)}  \\ \hline
    Input:$\bA,\bb, \eta>0,  S^0=\{i_1,\ldots,i_k\}$.\\
Iteration:  For $n=1,2,...,k$.\\
\quad 1) $\mbx^n= \hbox{arg}\,\,\min_{\bz}\|\bA \bz-\bb\|_2,\quad
 \hbox{supp}(\bz)=(S^{n-1} \backslash \{i_n\})\cup \{j_n\}, $ $  j_n\in B_\eta(\{i_n\})$.
\\
 \quad 2) $S^n=\hbox{supp}(\mbx^n)$.\\
    Output:  $S^k$.\\
    \hline
   \end{tabular}
\end{center}

\bigskip

We now give a condition under which LO does not spoil the
BOMP reconstruction$^8$. 
\begin{theorem}
\label{thm2}
Let $\eta>0$ and let $\mbx$ be a $s$-sparse vector such
that (\ref{sep}) holds. 
Let $S^0$ and $S^k$ be the input and output, respectively,  of the LO algorithm. 

If
\beq
x_{\rm min}> (\ep+2(s-1)\eta) \lt({1\over 1-\eta}+\sqrt{{1\over (1-\eta)^2}+{1\over 1-\eta^2}}\rt)\label{93}
\eeq
and each element of $S^0$  is in the $\eta$-coherence
band of a unique nonzero component of $\mbx$, then
 each element of $S^k$ remains in the $\eta$-coherence
band of a unique nonzero component of $\mbx$. 
\end{theorem}

Embedding LO in BOMP gives rise to the Band-excluded, Locally
Optimized Orthogonal Matching Pursuit (BLOOMP).

\bigskip

\begin{center}
   \begin{tabular}{l}
   \hline
   
   \centerline{{\bf Algorithm 3.} Band-excluded, Locally Optimized Orthogonal Matching Pursuit (BLOOMP)} \\ \hline
   Input: $\bA, \bb,\eta>0$\\
 Initialization:  $\mbx^0 = 0, \br^0 = \bb$ and $S^0=\emptyset$ \\ 
Iteration: For  $n=1,...,s$\\
\quad {1)  $i_{\rm max} = \hbox{arg}\max_{i}|\lan \br^{n-1},\ba_i\ran | , i \notin B^{(2)}_\eta(S^{n-1}) $} \\
  \quad      2) $S^{n} = \hbox{LO}(S^{n-1} \cup \{i_{\rm max}\})$ where $\hbox{LO}$ is the output of Algorithm 2.\\
  \quad  3) $\mbx^n = \hbox{arg}  \min_\bz \|
     \bA \bz-\bb\|_2$ s.t. \hbox{supp}($\bz$) $\in S^n$ \\
  \quad   4) $\br^n = \bb- \bA \mbx^n$\\
 Output: $\mbx^s$. \\
 \hline
   \end{tabular}
\end{center}

\bigskip

\begin{corollary}
\label{cor1}
Let $\hat \mbx$ be the output of BLOOMP. 
Under the assumptions of Theorems \ref{thm1} and \ref{thm2},
 $\hbox{supp}(\hat\mbx)\subseteq B_\eta(\hbox{supp}(\mbx))$
and moreover every nonzero component of $\hat\mbx$ is in
the $\eta$-coherence band of a unique nonzero component of $\mbx$. 
\end{corollary}
Even though we can not 
improve the performance guarantee for BLOOMP, in practice the LO technique 
greatly enhances the success probability of recovery  with respect to noise stability and dynamic range.  Moreover, if Corollary \ref{cor1} holds, then for all practical purposes
we have the residual  bound for the BLOOMP reconstruction $\hat\mbx$
\beq
\label{eb}
\|\bb-\bA\hat \mbx\|_2\leq c \|\mbe\|_2,\quad c\sim 1.
\eeq
  \\

\bigskip

\centerline{\bf \small 4. NUMERICAL RESULTS}

\bigskip

We test the  algorithms, BOMP and BLOOMP,  on  the two examples discussed in the Introduction.

For the first example (\ref{2})-(\ref{1}),  we use the refinement factor $F=20$. For the objects $\mbx$, we use 10 randomly phased and located  objects, separated by at least 3 $\ell_R$. 
The noise is  the i.i.d. Gaussian noise $\mbe\sim N(0,\sigma^2 I)$. 
 
 \begin{figure}
\centering
\includegraphics[width=8cm]{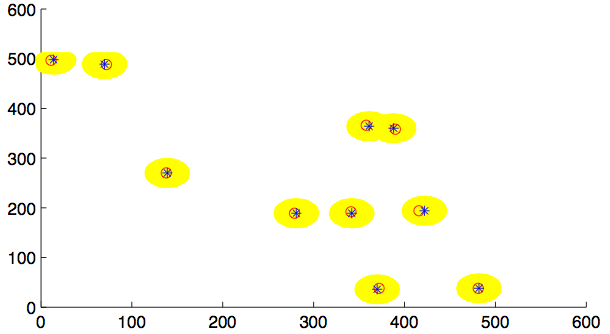}
\includegraphics[width=8cm]{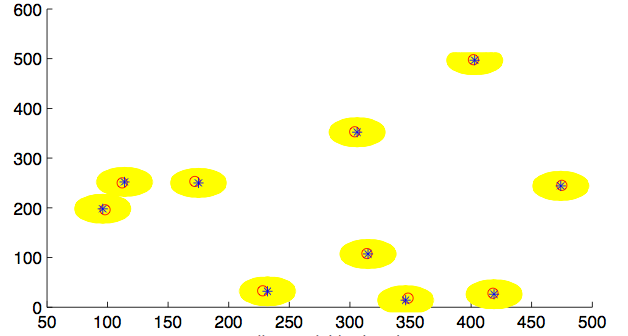}
\caption{Two instances of BOMP reconstruction: red circles are the exact locations, blue asterisks are recovered locations and the yellow
patches are the coherence bands around the objects.}
\label{fig-2d}
\end{figure}

Figure \ref{fig-2d} shows two instances of reconstruction by BOMP
in two dimensions.
The recovered objects (blue asterisks) are close to the true objects (red circles)
well within the coherence bands (yellow patches).
 
For the rest of simulations, we show the percentage of successes in
100 independent trials. A reconstruction is counted as a success  if every reconstructed  object
is within 1 $\ell_R$  of 
the object support. This is equivalent to
the criterion that the Bottleneck distance between
the true support and the reconstructed support
is less than 1 $\ell_R$. The result is shown in Figure \ref{fig5}.
With 10 objects of dynamic range $5$, BLOOMP requires the least number of
measurements, followed by BOMP
and then OMP, which does not  achieve high success rate even with 100 measurements (left panel). With 100 measurements ($N=100$) and 
$1\%$ noise,
BLOOMP can handle dynamic range up to 120
while BOMP and OMP  can handle dynamic range about 5 and 1, respectively.

\begin{figure}
\centering
\includegraphics[width=8cm]{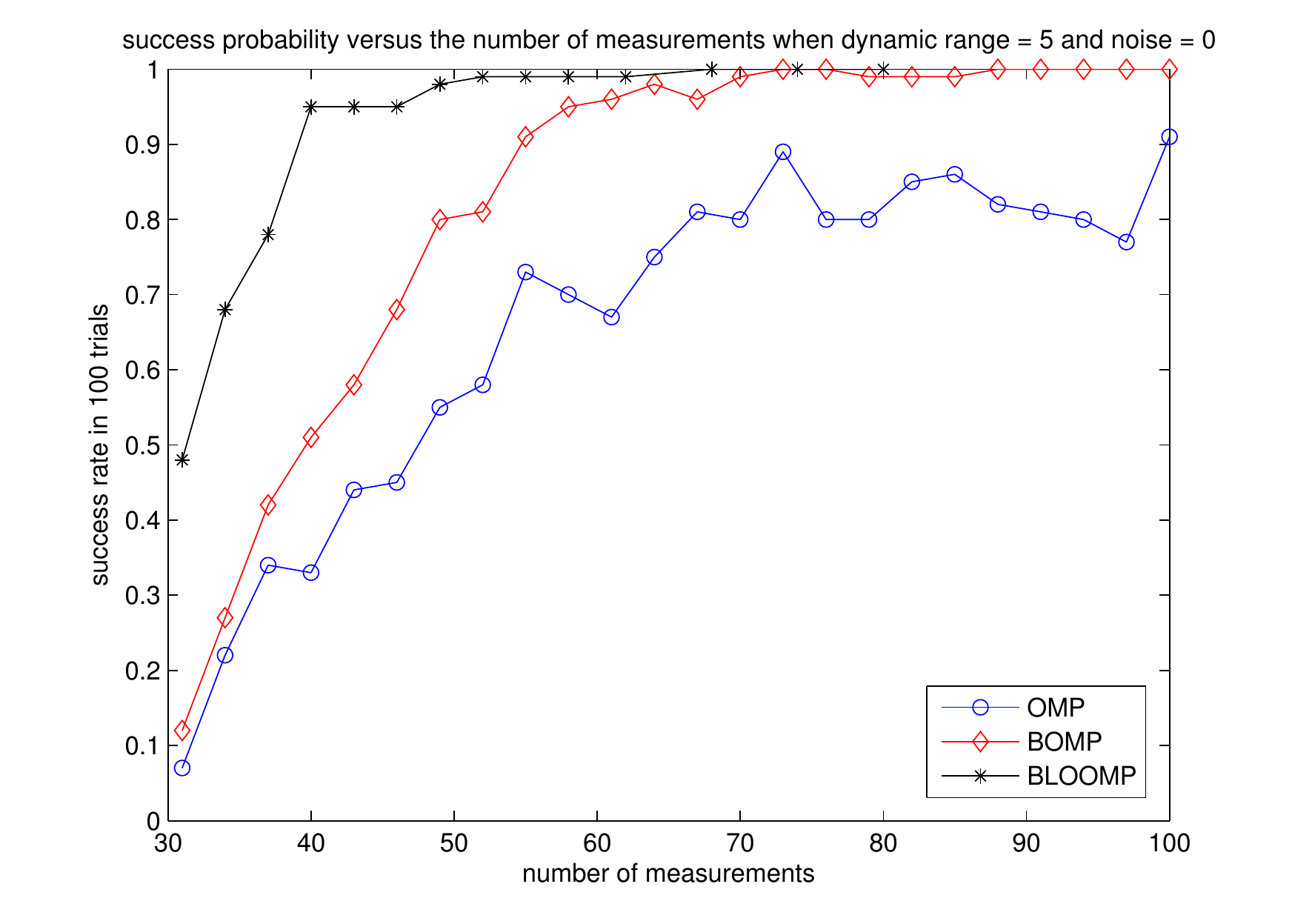}
\includegraphics[width=8cm]{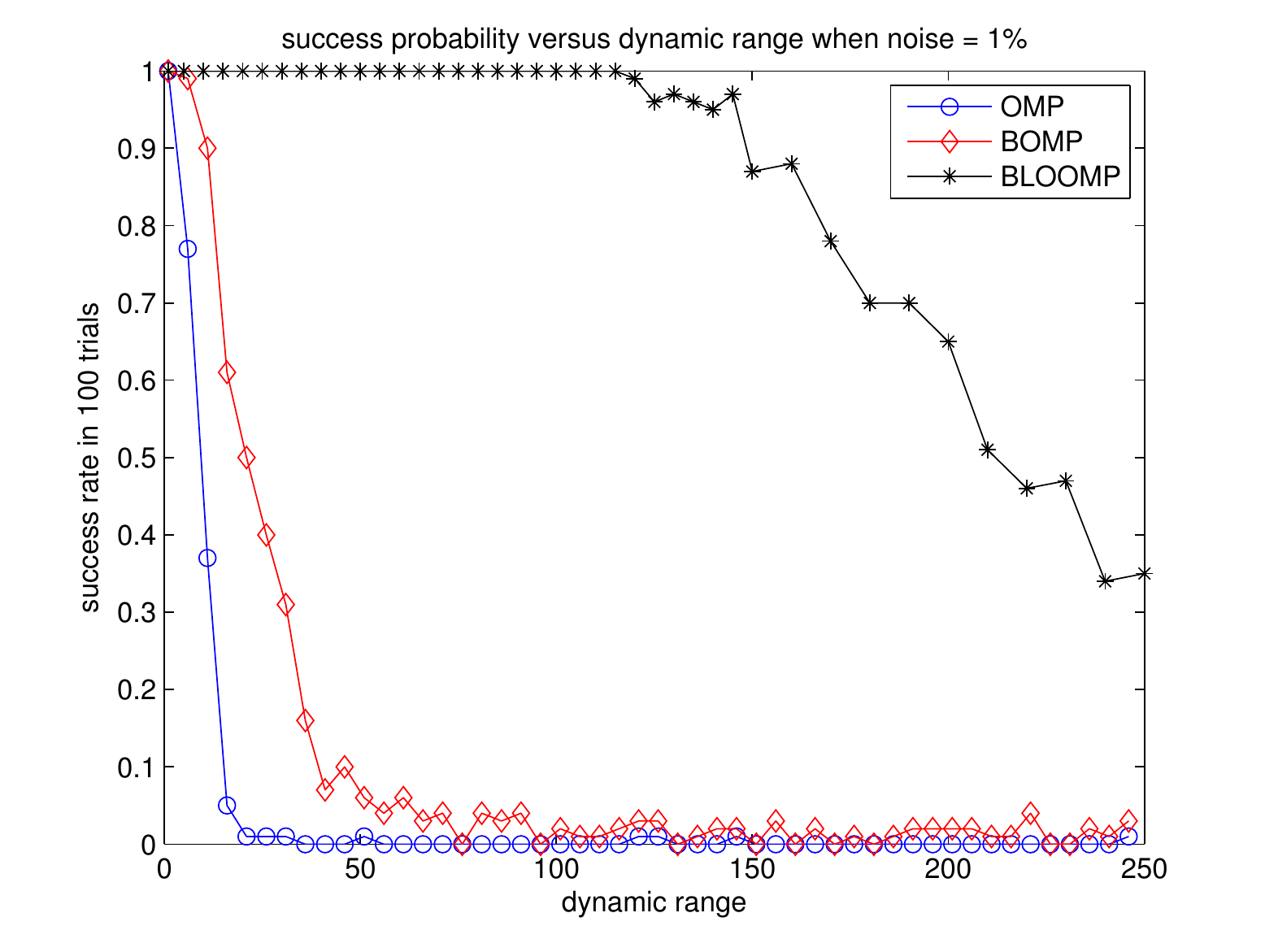}
\caption{Success rate versus number of measurements (left, dynamic range 5, zero noise) and dynamic range (right, $1\%$ noise) for OMP, BOMP and BLOOMP.}
\label{fig5}
\end{figure}

\begin{figure}
\centering
\includegraphics[width=8cm]{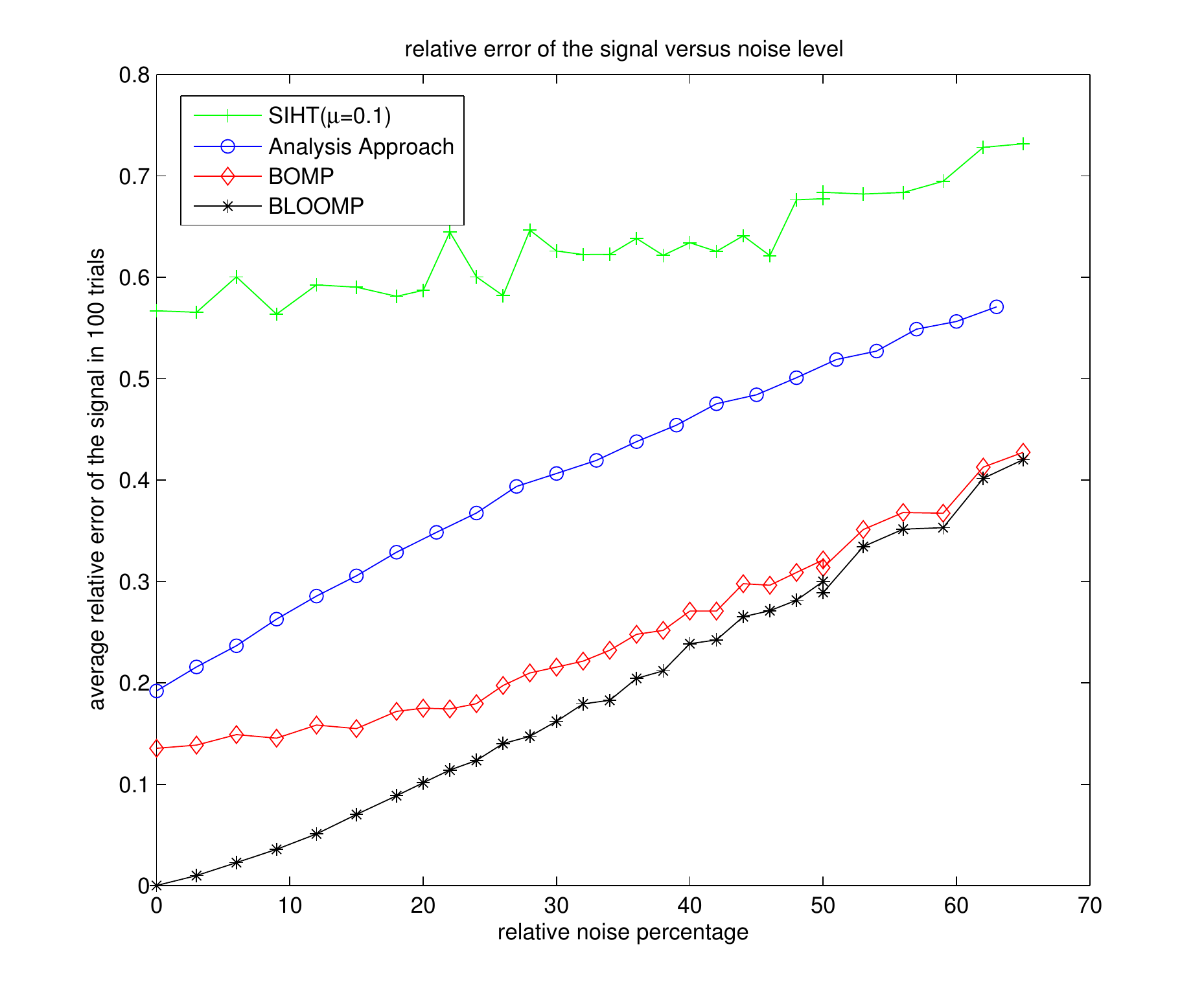}
\includegraphics[width=8cm]{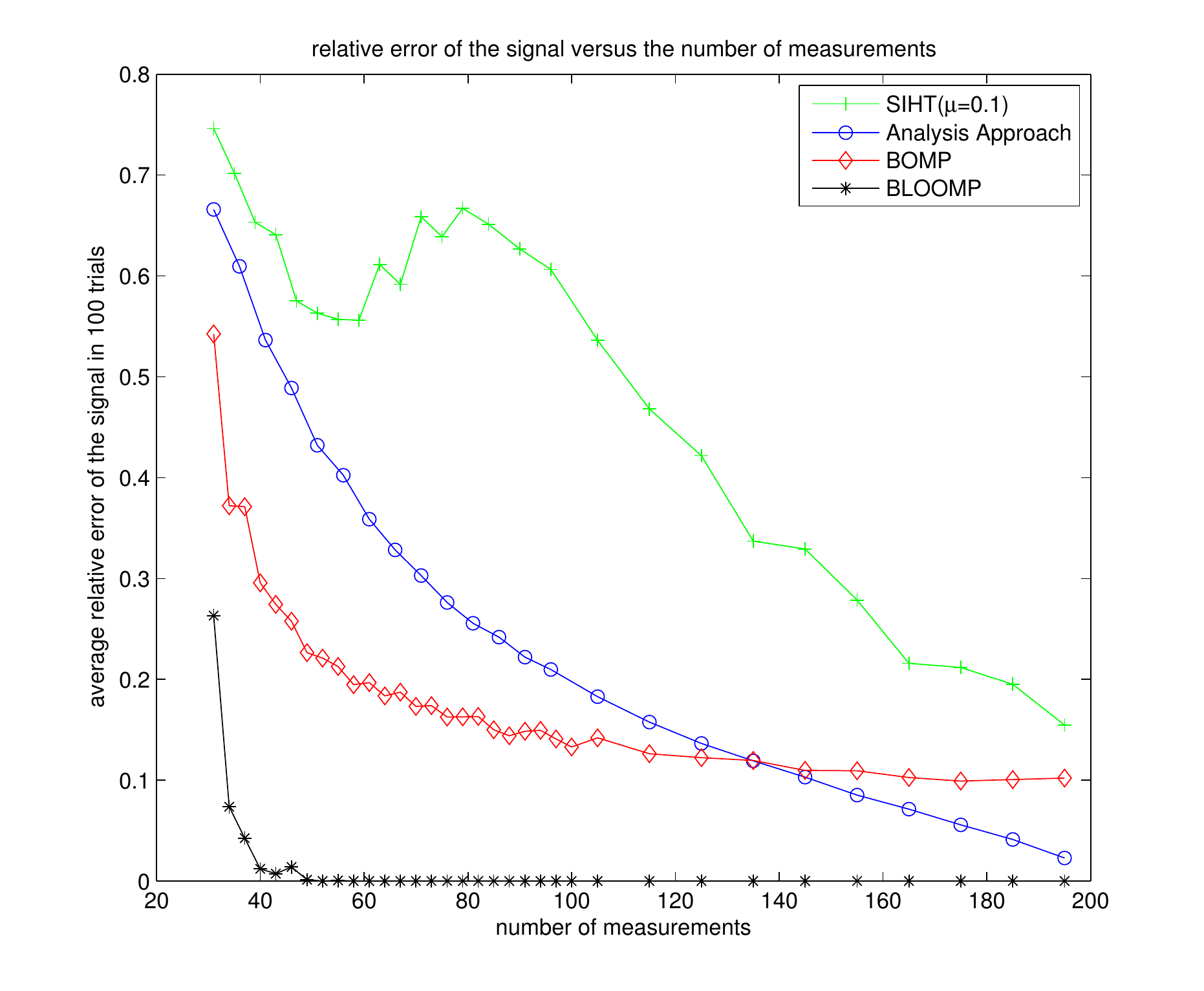}

\caption{Relative errors versus relative noise (left)  and number of measurements (right, zero noise) for dynamic range 10. }
\label{fig6}
\end{figure}

\commentout{
\begin{figure}
\centering
\includegraphics[width=10cm]{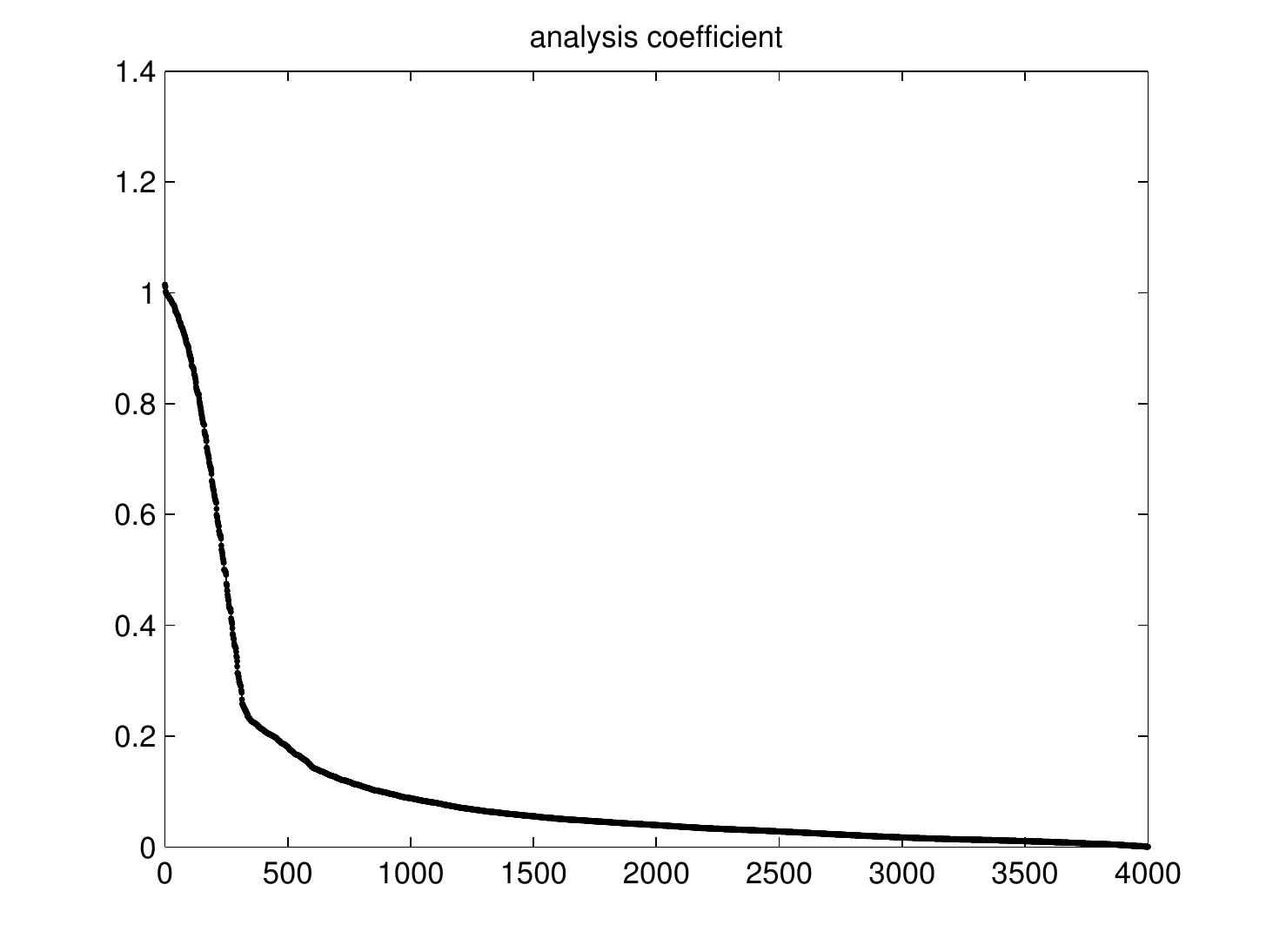}

\caption{Analysis coefficients $\bD^* \by$ reorganized according to magnitudes. The horizontal axis counts the number of coefficients greater than or equal to the height of the curve.} 
\label{fig7}
\end{figure}
}

For the second example (\ref{101})-(\ref{102}), we test, 
in addition to our algorithms, the method proposed 
by Duarte and Baraniuk$^5$ and 
the analysis approach of frame-adapted Basis Pursuit$^{2,6}$. 

The algorithm, Spectral Iterative Hard Thresholding (SIHT)$^5$,  assumes the model-based RIP which, in spirit, is equivalent 
  to the assumption of well separated support in the synthesis coefficients and therefore resembles  closely to our approach. 
  
While SIHT is a synthesis method like BOMP and  BLOOMP,  the frame-adapted BP  
 \beq
 \label{112}
 \text{min} \|\bPsi^{\star}\bz \|_1 \qquad \text{ s.t } \|\bPhi \bz -\bb \|_2 \le \|\mbe\|_2,
\eeq
is the analysis approach$^6$. 
Cand\`es {\em et al.}$^2$ have established a performance
guarantee for (\ref{112}) provided that  the measurement  matrix $\bPhi$ satisfies 
the frame-adapted  RIP:
\beq
\label{rip}
(1-\delta)\|\bPsi \bz\|_2 \leq\|\bPhi\bPsi \bz\|_2\leq (1+\delta)
\|\bPsi z\|_2,\quad \|\bz\|_0\leq 2s
\eeq
 for a tight frame $\bPsi$ and  a sufficiently small $\delta$ 
 and that  the analysis coefficients $\bPsi^*\by$
are sparse or compressible. 

\commentout{
Unfortunately, the latter condition may be easily  violated for signal $\by$
of the form $\by=\bPsi \mbx$ where $\mbx$ is $s$-sparse and widely separated. 
For example, Figure \ref{fig7}  shows the absolute values of
the component of the vector  $\bPsi^*\by$ in
the order of descending magnitude. Roughly speaking,  redundancy $F=20$ 
produces about $2F=40$ highly coherent columns around each synthesis coefficient and hence 
$\bPsi^*\by$ has about $400$ significant components. In general,
the sparsity of the analysis coefficients is at least $2sF$ where
$s$ is the sparsity of the synthesis coefficients and
$F$ is the redundancy. 
Therefore the analysis approach (\ref{112}) would  require  more than 100  
measurements   for accurate reconstruction and the result
 would depend on the redundancy. 
 }
 
 Instead of the synthesis coefficients $\mbx$, however,  the quantities  of interest are $\by$. Accordingly 
we measure
the performance  by the relative error $\|\hat\by -\by\|_2 / \|\by\|_2$ averaged over 100 independent trials. In each trial, 
10 randomly phased and located 
objects (i.e. $\mbx$) of dynamic range 10 and i.i.d. Gaussian $\bPhi$ are generated. We set $N=100, R = 200, F = 20$ for test of noise stability
and vary $N$ for test of measurement compression.

 As shown in Figure \ref{fig6}, BLOOMP is the best performer in
 noise stability (left panel)  and measurement compression (right panel). 
 BLOOMP
requires about 40 measurements to
achieve nearly perfect reconstruction while the other methods
require more than 200 measurements.   Despite the powerful error bound established in \cite{CEN}, the analysis approach (\ref{112}) needs more than 200 measurements for accurate recovery because the analysis coefficients $\bPsi^*\by$ are typically not sparse. Here redundancy $F=20$ 
produces about $2F=40$ highly coherent columns around each synthesis coefficient and hence 
$\bPsi^*\by$ has about $400$ significant components. In general,
the sparsity of the analysis coefficients is at least $2sF$ where
$s$ is the sparsity of the {\em widely separated synthesis} coefficients and
$F$ is the redundancy. 
Thus according to the error bound of \cite{CEN}  the performance of  the analysis approach (\ref{112}) 
 would degrade with the redundancy of the dictionary.

 To understand the superior performance of BLOOMP in this set-up let us give an error bound
using  (\ref{eb}) and (\ref{rip})
  \beq
 \|\bPsi (\mbx-\hat\mbx)\|_2 \leq{1\over 1-\delta} \|\bA(\mbx-\hat\mbx)\|_2\leq {1\over 1-\delta} \|\bb-\mbe- \bA\hat\mbx\|_2\leq {1+c\over 1-\delta}\|\mbe\|_2\label{21}
 \eeq
 where $\hat \mbx$ is the output of BLOOMP. This implies that 
 the reconstruction error of BLOOMP is essentially determined by
 the external noise, consistent with the left and right panels  of
 Figure \ref{fig6}, and is independent of the dictionary redundancy 
 if Corollary \ref{cor1} holds. 
In comparison, the BOMP result
 appears to approach an asymptote of  nonzero ($\sim 10\%$)  error. 
 This demonstrates the effect of local optimization technique in reducing
 error. The advantage of BLOOMP over BOMP, however,  disappears in the presence of large external noise (left panel).

\bigskip

\centerline{\bf \small 5. CONCLUSION}

\bigskip

We have proposed  algorithms, BOMP and BLOOMP,  for sparse recovery  with
highly coherent, redundant  sensing matrices and have established
performance guarantee that is {\em redundancy independent}. 
These algorithms have a sparsity constraint and computational
cost similar to OMP's. Our work is inspired by the
redundancy-independent performance guarantee recently established
for the MUSIC algorithm for array processing.$^7$

Our algorithms are based on variants of OMP enhanced by   two novel techniques: band exclusion and local optimization. We have extended these techniques to various CS algorithms, including Lasso,  and 
performed systematic tests elsewhere$^8$. 

Numerical results demonstrate the superiority of BLO-based  algorithms    for reconstruction of sparse objects separated by above the Rayleigh threshold. \\

\bigskip

\noindent{\bf Acknowledgements.} The research is partially supported in part by NSF Grant DMS 0908535. 

\bigskip

\bibliographystyle{plain}	
\bibliography{myref}		
 
\end{document}